\documentclass{WileyMSP-template}
\usepackage{amsmath}
\usepackage{hyperref}
\hypersetup{
colorlinks=true,
linkcolor=blue,
filecolor=magenta,
urlcolor=blue,
}
\usepackage{pdfpages}
\begin{document}

\pagestyle{fancy}
\rhead{Impact of boron atom clustering on the electronic structure of (B,In)N alloys}

\title{Impact of boron atom clustering on the electronic structure of (B,In)N alloys}

\maketitle


\author{Cara-Lena Nies*}
\author{Stefan Schulz*}



\begin{affiliations}
Dr Cara-Lena Nies, Dr Stefan Schulz\\
Tyndall National Institute, 
University College Cork, Lee Maltings, Dyke Parade, 
Cork, Ireland\\
Email Address: caralena.nies@tyndall.ie; stefan.schulz@tyndall.ie\\

Dr Stefan Schulz\\
School of Physics, 
University College Cork, College Road, 
Cork, Ireland\\

\end{affiliations}


\keywords{III-nitrides, DFT, alloy microstructure}

\begin{abstract}

Tailoring the electronic and optical properties of nitride-based alloys for optoelectronic applications in the ultraviolet and red spectral range has attracted significant attention in recent years. Adding boron nitride (BN) to indium gallium nitride, (In,Ga)N, alloys, can help to control the lattice mismatch between (In,Ga)N and GaN and may thus allow reduction of strain related defect formation. However, understanding of the impact of BN on the electronic properties of III-N alloys, in particular the influence of experimentally observed boron atom clustering, is sparse. This work presents first-principles calculations investigating the electronic properties of (B,In)N alloys with boron contents between 2\% and 7\%. Special attention is paid to the impact of the alloy microstructure. While the results show that the lattice constants of such alloys largely agree with lattice constants determined from a Vegard approximation, the electronic properties strongly depend on the local boron atom configurations. For instance, if boron atoms are dispersed throughout the structure and are not sharing nitrogen atoms, the band gap of (B,In)N alloys is largely unaffected and stays close to the gap of pristine InN. However, in the case of boron atom clustering, e.g., when boron atoms are sharing nitrogen atoms, the band gap can be strongly reduced, often leading to a metallic state in (B,In)N alloys. These strong band gap reductions are mainly driven by carrier localization effects in the valence band. The calculations thus show that the electronic structure of (B,In)N alloys strongly depends on the alloy microstructure and that boron atom clustering plays an important role in understanding the electronic and optical properties of these emerging materials. 

\end{abstract}


\section{Introduction}

The low quantum efficiency of deep UV and red light emitting diodes (LEDs) based on III-nitride (III-N) materials is a well known bottleneck towards creating efficient, commercially available LEDs in the 220-280 nm and 620-750 nm wavelength ranges~\cite{amano2020, iida21}. Efficient red LEDs are of particular interest for creating full color LED displays, including micro-LEDs for augmented (AR) and virtual reality (VR) applications~\cite{iida21, wierer19, jiang21}. In principle, (In,Ga)N based LEDs can provide all the LEDs required for these displays as the band gap of such alloys can span the entire visible spectrum. Particularly devices for shorter wavelengths (blue-violet) exhibit high quantum efficiencies~\cite{iida21}. While it is possible to create red (In,Ga)N LEDs~\cite{hwang14, dussaigne20, chan21}, an increased In content compared to blue and green LEDs is required to achieve these longer wavelengths~\cite{iida21}. Increasing the In content requires different growth conditions and also increases the lattice mismatch in the quantum well heterostructures, leading to a rise in defect density which contributes to poor efficiencies~\cite{chan21,oliver05}. For this reason, the red component of full color LED displays is currently made utilizing (Al,Ga,In)P alloys, which demonstrate external quantum efficiencies exceeding 60\%~\cite{broell14}. However, being able to create full color LED displays but also micro-LEDs for AR and VR applications based on a single material system has many advantages for the manufacturing process, creating significant interest in improving the quality of (In,Ga)N based light emitters~\cite{iida21, jiang21}. Additionally, (In,Ga)N micro-LEDs promise improved efficiency compared to (Al,Ga,In)P based micro-LEDs of similar dimensions, due to the higher surface recombination and increased carrier lifetime of (Al,Ga,In)P~\cite{iida21, bulashevich16}. 

Alloying III-N materials with boron has recently emerged as a potential method for combating the lattice mismatch in III-N alloys and thus improve material quality and quantum efficiency~\cite{williams17, gunning17, williams19, kudrawiec20, mickevivcius19}. While the lowest energy polymorph of BN has a hexagonal crystal structure, BN also exists in the wurtzite phase, making it suitable for alloying with the other III-N materials, AlN, GaN and InN, which preferentially grow in the wurtzite (wz) crystal structure. Due to the inherent size difference, the much smaller lattice parameter of wz-BN (e.g., the lattice mismatch between BN and InN is approximately 30\%) can aid in strain-compensation of III-N heterostructures. This should ideally reduce both the strain-induced defects as well as strain-induced built-in polarization fields~\cite{minj13, caro13}. However, unlike AlN, GaN and InN, wz-BN is an \emph{indirect} band gap material with a gap of around 6.7 eV from $\Gamma$ to K~\cite{BGaN,sheerin22}. Introduction of a material with a very different electronic structure but also structural properties (e.g. atomic size) can have unforeseen effects on the material properties as is reported for many highly mismatched alloys~\cite{walukiewic20,OReLi2009,ScCa2014}. In order for boron incorporation to become an effective method to alleviate lattice mismatch in e.g. (B,In,Ga)N/GaN heterostructures, the evolution of the band gap of such alloys should ideally be dominated by the direct gap materials, e.g. InN and GaN. Otherwise, boron-induced changes to the electronic structure should be harnessed to tune the emission wavelength of the material. Initial explorations of the fundamental properties of (B,Ga)N~\cite{turiansky19, BGaN} and (B,In,Ga)N~\cite{williams17} alloys from first principles calculations have already been reported in the literature. This includes our recent work on the influence of the alloy microstructure, in particular boron atom clustering, on the electronic structure of (B,Ga)N alloys~\cite{BGaN}, an effect widely overlooked so far in the literature (for instance Refs.~\cite{turiansky19} and~\cite{williams17} assume a random alloy). Our calculations on (B,Ga)N alloys show that boron atom clusters can significantly impact the electronic structure and lead to carrier localization effects and accompanying strong band gap reductions. This highlights that the local alloy composition may have a significant influence on the electronic structure but also the optical properties of boron containing III-N alloys. 

As indicated above, (B,In,Ga)N alloys are candidate materials to design efficient light emitters in the red spectral region. However, given that the alloy microstructure in both (In,Ga)N~\cite{ScCa2015} and (B,Ga)N~\cite{BGaN} alloys already plays an important role for their optoelectronic properties, one may expect that the electronic structure of (B,In,Ga)N alloys is strongly impacted by alloy disorder. In order to gain a more in-depth understanding of the electronic properties of (B,In,Ga)N alloys for future applications, it is useful to analyze the impact of wz BN on the electronic structure in the absence of GaN to provide a basis for future studies of the more complex (B,In,Ga)N alloy systems. This study therefore aims to understand the electronic properties of (B,In)N alloys. While this material system is very challenging to grow due to the extremely large lattice mismatch between wz BN and InN, a theoretical investigation is nonetheless valuable and can then be compared to our study of (B,Ga)N alloys. We note that while \emph{hexagonal} (B,In)N has previously been studied in the literature~\cite{chen09}, to the best of our knowledge there is no such investigation for \emph{wz} (B,In)N. In the following, we target the different boron microstructures investigated in our previous work on (B,Ga)N, allowing us to analyse the impact of boron atom clustering on the electronic structure of (B,In)N alloys.

\section{Methods}
We model (B,In)N alloys atomistically using plane-wave density functional theory (DFT) as implemented in the Vienna ab initio Simulation Package (VASP) v.5.4.4~\cite{vasp}. The atomistic structures and charge densities were visualized using VESTA~\cite{VESTA}. For the geometry and lattice optimizations we used the generalized gradient approximation (GGA) based on the the Perdew-Burke-Ernzerhof (PBE)~\cite{PBE} approach to the exchange correlation functional. All electronic structure calculations were carried out using the modified Becke-Johnson (mBJ)~\cite{BJ,mBJ} meta-GGA functional, benchmarked to state of the art hybrid DFT calculations using the Heyd-Scuseria-Ernzerhof (HSE)~\cite{HSE,HSE06} functional, ensuring that that the small bang gap of InN (0.69 eV) is captured accurately. Benchmarks, band structures and mBJ parameters ($c$-parameters) can be found in our previous work~\cite{BGaN}. For both calculations using PBE and mBJ, we employ the In pseudopotentials which describe the semi-core $d$-electrons as valence electrons. This means that for our system the valence electron configurations are as follow: B 2s$^2$2p$^1$, In 5s$^2$4d$^10$5p$^1$, N 2s$^2$2p$^3$. As in our previous work on (B,Ga)N alloys, we use a $3\times3\times3$ supercell with 108 atoms in which 1 to 4 In atoms are replaced with B atoms, equivalent to approximately 2-7\% B content. A $\Gamma$-centered $2\times2\times1$ Monkhorst Pack k-point mesh is employed and all calculations are performed with a plane wave cut-off energy of 600 eV. Density of state (DOS) calculations were scaled such that the energy of the highest occupied state at the $\Gamma$-point is equal to 0, as both wz BN and InN have their valence band maximum at $\Gamma$. For the majority of (B,In)N alloy configurations, we employ a Gaussian smearing of $\sigma$ = 0.01 eV. Due to the small band gap of InN (0.69 eV), this can sometimes lead to partially occupied states around the band gap ($\Gamma$-point) in which case the smearing has been decreased to $\sigma$ = 0.0001 eV for the band structure calculation. Tests using InN showed that this change in smearing does not affect the magnitude of the band gap. The lattice parameters are optimized such that the pressure of the supercell is minimized, see Supporting Information. Lattice parameters obtained via this method do not vary from those predicted based on Vegard's approximation by more than $\pm$ 2\% (see also discussion below).

\section{Results and Discussion}
In this section we present results of our first-principles electronic structure calculations for (B,In)N alloys. Section~\ref{subsec:alloyconfig} gives a brief overview of the different alloy configurations studied. In Section~\ref{subsec:EgEvolve}, we discuss how the different alloy configurations change the band gap of (B,In)N alloys. 

\subsection{Alloy configurations in (B,In)N alloys}
\label{subsec:alloyconfig}

As already mentioned above, for this work we used the same boron atom cluster configurations that we have previously studied for (B,Ga)N alloys~\cite{BGaN}; visual representations of the different configurations are given in the Supporting Information of Ref.~\cite{BGaN}. In general we study here:
\begin{itemize}
    \item \textbf{Close:} 3 or 4 B atoms substituted at cation sites around a single N atom, i.e. in case of 4 B atoms a full tetrahedron is formed.
    \item \textbf{Apart:} 2, 3 or 4 B atoms substituted with the maximum number of cation sites between them so that they are not sharing a N atom.
    \item \textbf{Line:} 2, 3 or 4 B atoms substituted at adjacent cation sites always connected by a N atom in the same $c$-\emph{plane}. 
    \item \textbf{z-Line:} 2, 3 or 4 B atoms substituted along the $c$-\emph{axis} (i.e., one, one and a half or two unit cells of BN stacked along the wz $c$-axis).
\end{itemize}

\subsection{Band gap Evolution}
\label{subsec:EgEvolve}

\begin{figure}
\begin{tabular}{|c|c|}
\hline
\textbf{(a) Band Gap Evolution} & \textbf{(b) Lattice Parameters}\\\hline 
\includegraphics[width=0.475\columnwidth]{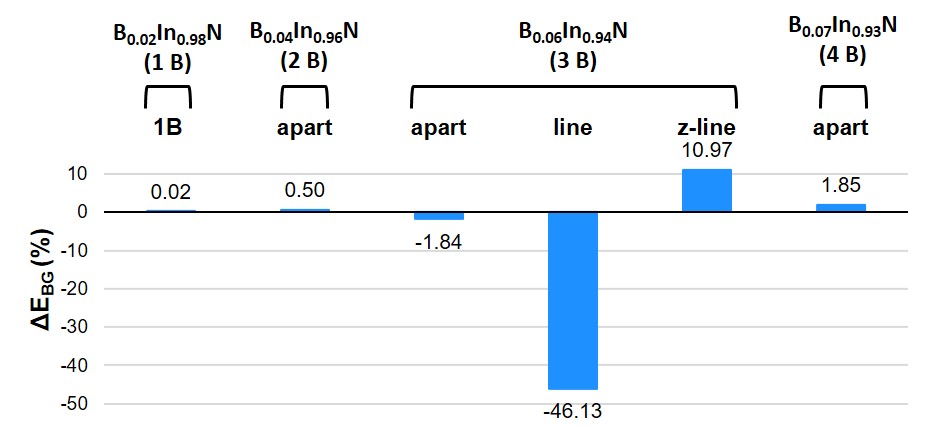} & \includegraphics[width=0.4755\columnwidth]{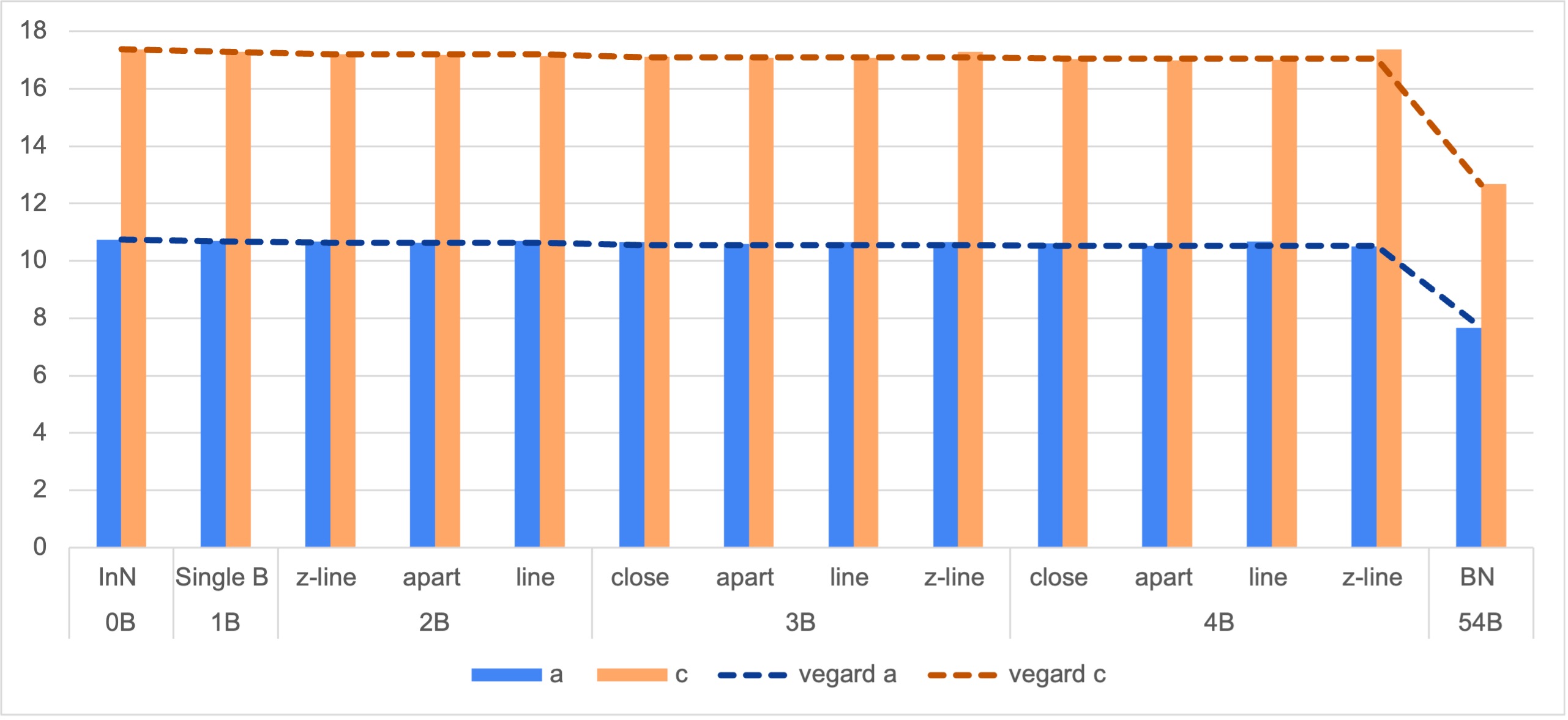}\\\hline
\end{tabular}
\caption{(a) Percentage change in band gap of (B,In)N alloys with respect to InN. For the definition of the different alloy configurations, see main text. (b) Optimized lattice parameters for different alloy configurations, see main text, in the 108 atom supercell (54 cations, 54 nitrogen atoms) underlying our first-principles calculations. The (in-plane) $a$-lattice constant of the 108 atom supercell is shown in blue, while the $c$-lattice constant (along the wurtzite $c$-axis) is shown in orange. The dashed lines indicate the lattice constants predicted from a Vegard approximation for a B$_x$In$_{1-x}$ alloy: $a_\text{BInN}=xa_\text{BN}+(1-x)a_\text{InN}$; $c_\text{BInN}=xc_\text{BN}+(1-x)c_\text{InN}$.}
\label{fig:BG-lat}
\end{figure}

\begin{figure}[t!]
    \centering
    \includegraphics[width=\linewidth,keepaspectratio]{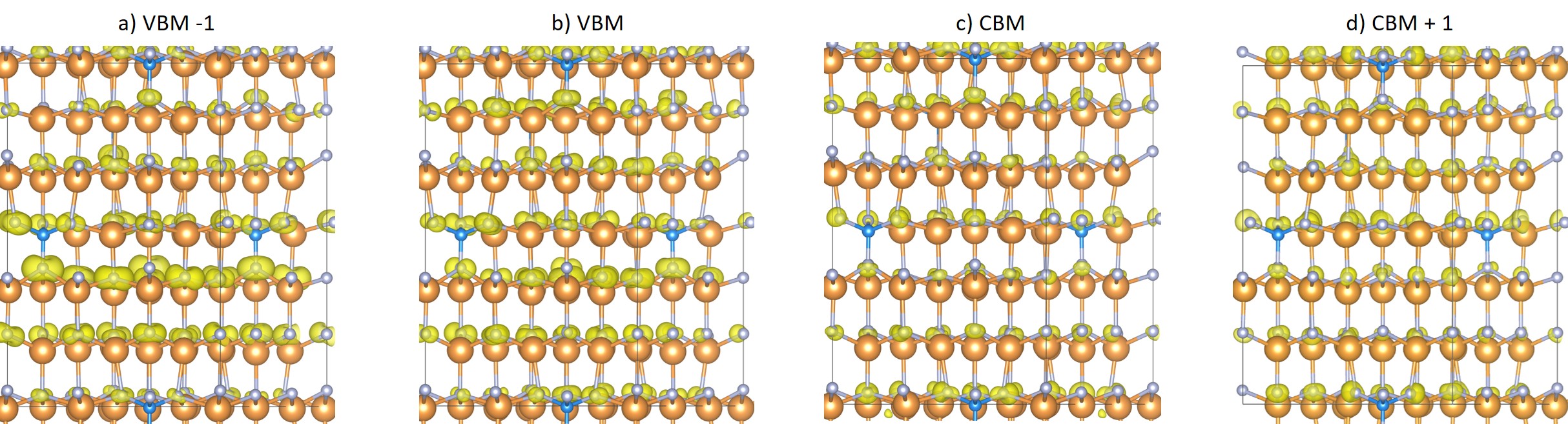}
    \caption{Band decomposed charge densities for a 4 B atoms distributed in InN. In = orange spheres, B = blue spheres, N = silver spheres, charge density = yellow areas. The charge densities are plotted using a constant isosurface level of 0.0013 in VESTA. Please note that any potential gaps in the crystal structure are caused by visualising a system with periodic boundary conditions.}
    \label{fig:4B-apart}
\end{figure}

\textbf{Figure~\ref{fig:BG-lat} (a)} summarises the change in band gap relative to InN for several of the studied alloy configurations. For the configurations in which the B atoms are distributed throughout the supercell (labelled "apart"), so that they are not sharing N atoms and are as far apart as possible considering the periodic boundary conditions, we find that the band gap changes very little. Similarly, the lattice constant changes close to Vegard's approximation, as illustrated by \textbf{Figure~\ref{fig:BG-lat} (b)}. The exact lattice parameters for each configuration are included in the Supporting Information. The weak impact of distributed B atoms on the electronic structure is also reflected in the charge densities of the first two states energetically closest to the conduction band minimum (CBM and CBM+1) and valence band maximum (VBM and VBM-1) shown in \textbf{Figure~\ref{fig:4B-apart}} for the 4 B "apart" configuration. This delocalized nature of the charge densities is also found for "apart" configurations with a lower number of B atoms (1, 2 and 3). The weak impact of distributed, non-clustered B atoms on the electronic structure accompanied by a Vegard-like lattice constant evolution in (B,In)N is similar to our previous results for (B,Ga)N under the same alloy conditions~\cite{BGaN}. This indicates that if such an alloy configuration (B atoms not sharing N atoms) can be achieved experimentally in (B,In,Ga)N alloys, one could indeed tailor the lattice constant and thus control the strain in a (B,In,Ga)N heterostructure while the band gap evolution would be mainly determined by the In content of the alloy. Therefore, this could make (B,In,Ga)N alloys highly attractive to achieve efficient light emission in the red spectral range.  

\begin{figure}
\begin{tabular}{|c|c|}
\hline
\textbf{(a) 2B-InN apart DOS} & \textbf{(b) 2B-InN line DOS}\\\hline 
\includegraphics[width=0.475\columnwidth]{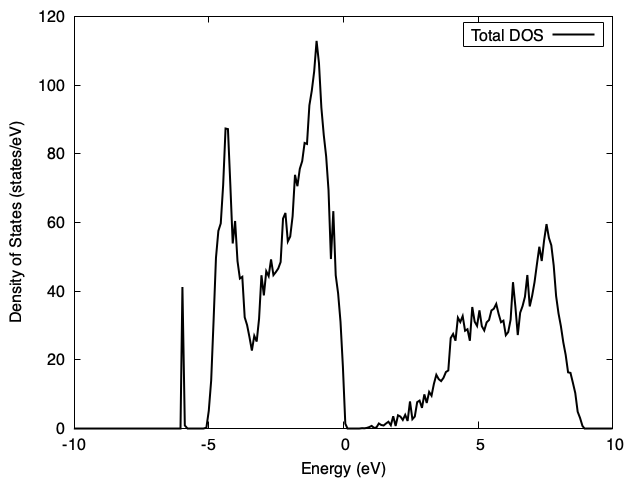} & \includegraphics[width=0.4755\columnwidth]{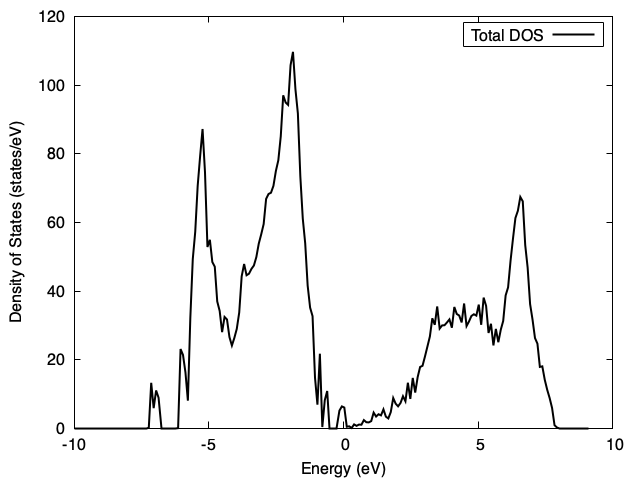}\\\hline
\end{tabular}
\caption{Density of States for (a) the 2 B atom "apart" configuration which indicates a semiconducting behavior and (b) 2 B "line" arrangement which exhibits metallic states. Details on the presented alloy configurations are given in the main text.}
\label{fig:DOS}
\end{figure}

However, our calculations reveal that when B atoms occupy adjacent cation sites in InN, thus sharing N atoms, the electronic structure of (B,In)N alloys is significantly affected. In general it is of note that the structural properties of InN and wz BN are very different due to the large size difference between the ionic radii of B (0.11~\AA) and In (0.62~\AA)~\cite{radii} and the lattice mismatch of approximately 30\% between wz BN and InN~\cite{BGaN}. This gives rise to very large local strain and polarization field effects, which can result in strong carrier effects, as we have already observed in (B,Ga)N alloys~\cite{BGaN}. In a variety of different highly mismatched alloys, localized states can lead to a strong and potentially composition dependent band gap reduction~\cite{OReLi2009,ScCa2013Apex}. Given the small band gap of InN, we observe that many alloy configurations involving B atom clustering show metallic behavior. To check if a metallic or semiconducting state is formed, we have not only calculated the eigenvalues and occupancies at the $\Gamma$-point of the supercell, but also at several points in k-space surrounding the $\Gamma$-point. This allows us to gain insight into the dispersion of the highest occupied and lowest unoccupied bands, which should form the valence and conduction band edges, respectively. Given that wz InN and BN have their VBM at $\Gamma$, we expect that in the semiconducting phase the energy of the valence band edge decreases away from $\Gamma$. However, if for instance for the "valence band edge" the energy eigenvalues increase when moving away from $\Gamma$ and the occupancy of the states is non-zero, the "opposite" band dispersion indicates a metal. This analysis is further corroborated by calculating the DOS, which reveals the formation of "defect bands" in the case of configurations identified as metallic by the band edge dispersion, see examples in \textbf{Figure~\ref{fig:DOS}} and Supporting Information. Our results highlight that it is essential to carry out such additional analysis when studying the band gap of a "narrow gap" semiconductor like InN, as the eigenvalues and occupancies at $\Gamma$ might otherwise suggest e.g., a widening of the band gap.

\textbf{Figure~\ref{fig:BG-lat} (a)} shows that out of all the configurations with boron atom clustering, only the 3 B atom configurations ("line" and "z-line") still exhibit a band gap. To understand the origin of this behavior, we will first discuss the "line" configuration, which corresponds to a one dimensional B-N-B chain in the wz $c$-plane. Unlike the equivalent 2 B and 4 B configurations, the 3 B configuration is semiconducting, however, it exhibits a band gap that is approximately 50\% smaller (0.36 eV) than the InN gap (0.69 eV). At first glance, there is no obvious origin for this difference. In fact, given that the 3 B "line" is semiconducting, one would expect that the 2 B configuration is also semiconducting due to its lower B content. However, our results indicate that the band gap is affected by more than just the boron content and the proximity of B atoms. A closer analysis of local bond length and angle changes, and thus strain fields, shows that the presence of boron can significantly distort the crystal structure. We first measure the interlayer distance, which we are taking to be the distance between N and B/In directly above each other along the $c$-axis, i.e. the cation-N bond length along the $c$-axis. This shows that the interlayer distance for the boron-containing layer is the same in all three configurations. It has shrunk compared to the InN interlayer distance and is now the same as in BN. The distance between the remaining InN layers does not change significantly from the distance in InN to compensate this change. In contrast, the N-B-N angle is not independent of the number of B atoms in the "line" configuration. For the metallic "line" configurations, this angle deviates significantly from the ideal tetrahedral bond angle in \emph{wz} BN, which is approximately 109$^{\circ}$. The 2 B "line" configuration has an N-B-N angle of 121$^{\circ}$, the 4 B "line" configuration has N-B-N angles ranging between 109$^{\circ}$ to 158$^{\circ}$, while the N-B-N angles for the semiconducting 3 B "line" configuration are close to wz BN with 107$^{\circ}$. In fact, due to the periodic boundary conditions, the 3 B "line" corresponds to a thin BN "wire" lying in the $c$-plane, which does not allow the B atoms to deviate significantly from the atomic position within the $c$-plane. 
\begin{figure}[t!]
    \centering
    \includegraphics[width=\linewidth,keepaspectratio]{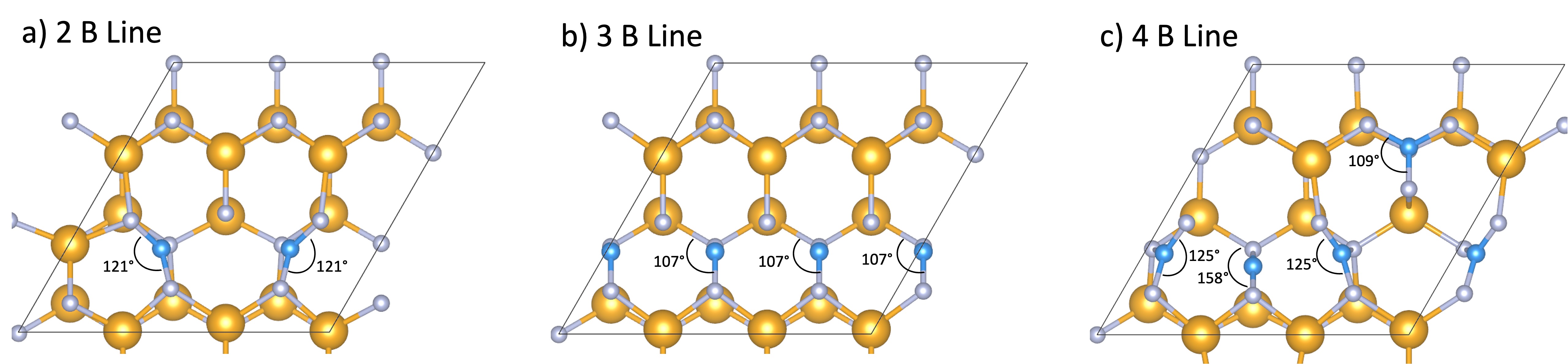}
    \caption{N-B-N angles in 2, 3 and 4 B atom "line" configurations. B configurations are shown looking down the $c$-axis with the bulk InN layers removed for easier viewing. See previous figure caption for information on color coding.}
    \label{fig:angles}
\end{figure}

\begin{figure}[ht!]
    \centering
    \includegraphics[width=\linewidth,keepaspectratio]{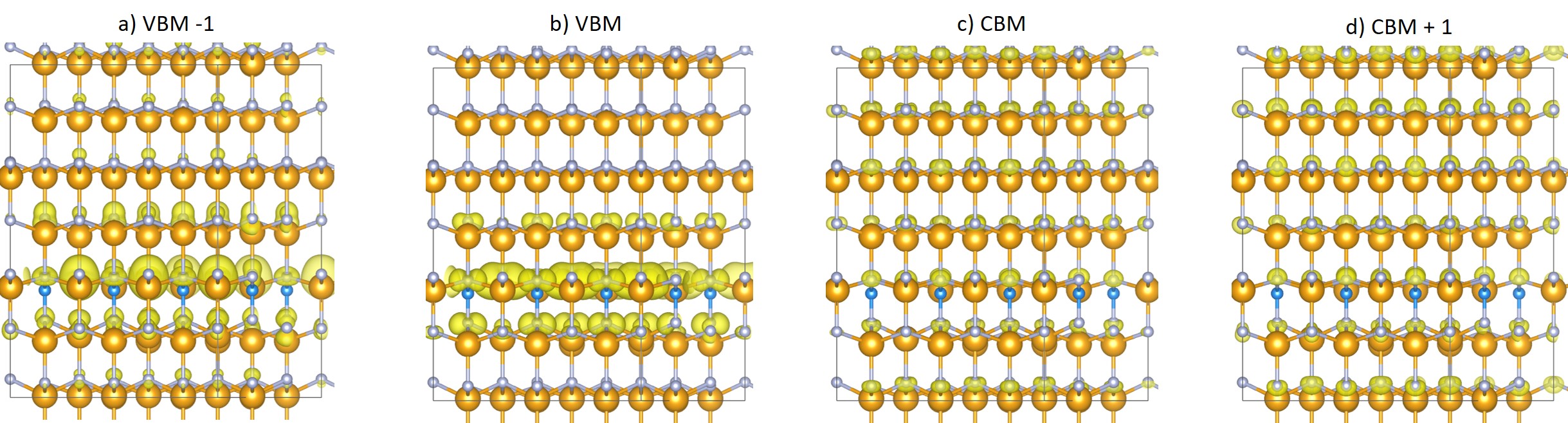}
    \caption{Band decomposed charge densities for a 3 B atoms in a horizontal "line" in InN. Note that the added number of B atoms visible in the image is due to the periodic boundary of the model. See previous figure caption for information on color coding.}
    \label{fig:3B-line}
\end{figure}

\begin{figure}[ht!]
    \centering
    \includegraphics[width=\linewidth,keepaspectratio]{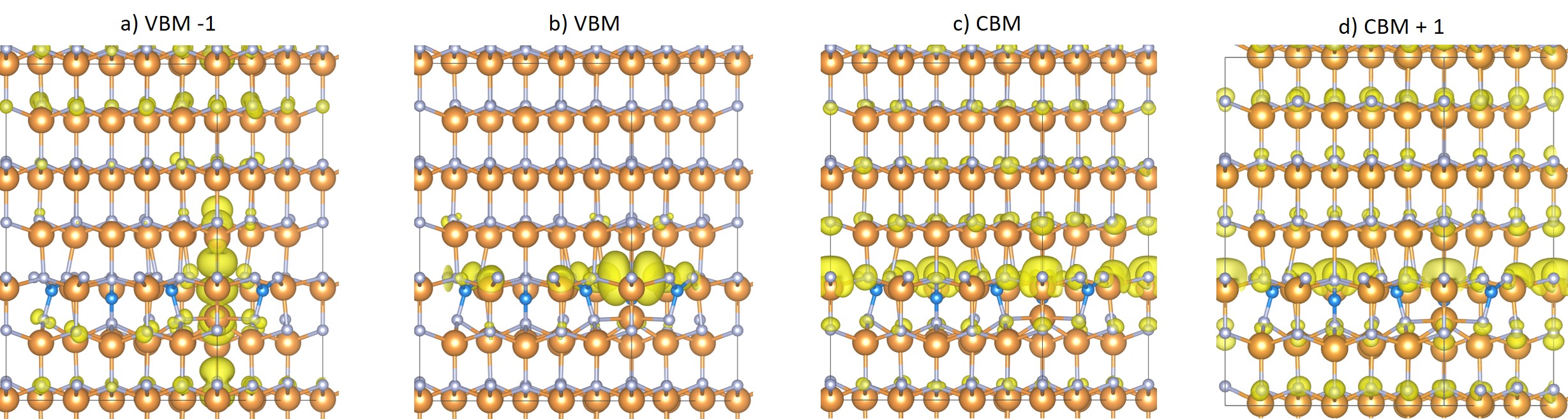}
    \caption{Band decomposed charge densities for a 4 B atoms in a horizontal "line" in InN. Note that the added number of B atoms visible in the image is due to the periodic boundary of the model. See previous figure caption for information on color coding.}
    \label{fig:4B-line}
\end{figure}
\textbf{Figure~\ref{fig:angles}} shows the relaxed atomic positions of all three configurations for a view along the wz $c$-axis and indicates the relevant N-B-N angles. The changes in bond angle around the B atoms in the 2 B and 4 B "line" configurations indicates a preference for B atoms in the same layer to relax into geometries closer to \emph{hexagonal} BN, where the same N-B-N angle is 120$^{\circ}$, i.e. the base of the tetrahedron for B atoms in this particular configuration is "flatter" and B and N atoms begin to share a layer rather than occupy distinct layers as is normal in wurtzite. Figure~\ref{fig:angles} also shows clearly that 2 B and 4 B "line" configurations cause distortions to the surrounding InN matrix, suggesting that both configurations cause large local strain. This is further confirmed by the band decomposed charge densities in \textbf{Figures~\ref{fig:3B-line}} and \textbf{\ref{fig:4B-line}}. These indicate that the carrier localization in the 4 B atom configuration is much stronger and occurs in both conduction and valence band levels in contrast to the 3 B configuration, which shows weaker localization effects mainly occurring in the VBM. This effect is also reflected in the densities of state shown in the Supporting Information, which indicate that a defect band forms in the band gap of InN in the presence of 2 B and 4 B "line" configurations. No such states are formed for the 3 B "line" configuration. 

\begin{figure}[ht!]
    \centering
    \includegraphics[width=\linewidth,keepaspectratio]{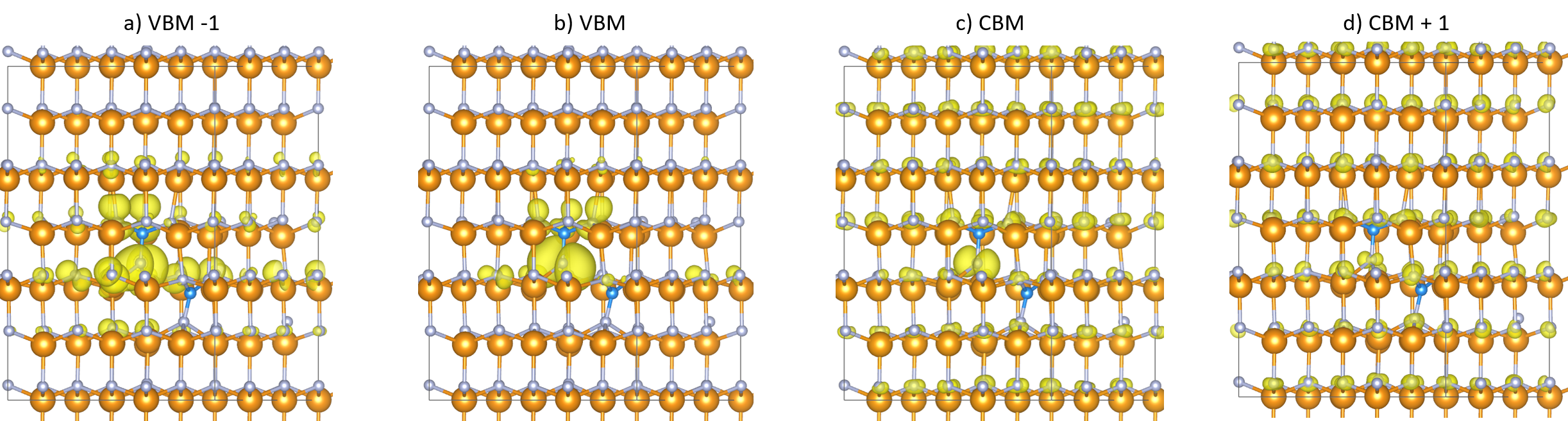}
    \caption{Band decomposed charge densities for a 2 B atoms in a vertical "z-line" in InN. See previous figure caption for information on color coding.}
    \label{fig:2B-zline}
\end{figure}

\begin{figure}[ht!]
    \centering
    \includegraphics[width=\linewidth,keepaspectratio]{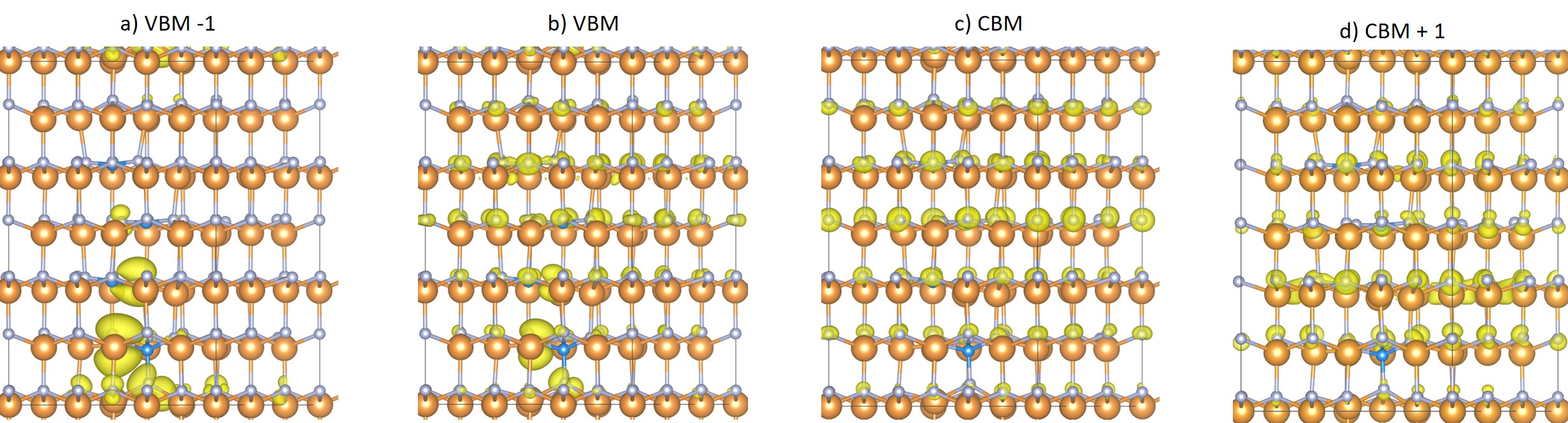}
    \caption{Band decomposed charge densities for a 4 B atoms in a vertical "z-line" in InN. See previous figure caption for information on color coding.}
    \label{fig:4B-zline}
\end{figure}

The \emph{3 B "z-line"} configuration is the only other clustered configuration which retains a band gap; the 2 B and 4 B atom "z-line" configurations are metallic. However, unlike the 3 B "line" configuration, the band gap of the "z-line" configuration increases slightly by 0.07 eV with respect to the InN band gap. To shed more light onto this behavior, we study the geometries and the band decomposed charge densities for the "z-line" configurations with 2 B (cf. \textbf{Figure~\ref{fig:2B-zline}}), 3 B (cf. \textbf{Figure~\ref{fig:3B-zline}}) and 4 B atoms (cf. \textbf{Figure~\ref{fig:4B-zline}}). In general we observe that bond lengths around the B atoms and thus also neighboring In atoms are very different in the different configurations studied here. For instance, in the 2 B case, Figure~\ref{fig:2B-zline}, the B atoms are significantly displaced from lattice positions of an ideal wz InN lattice. This is also reflected in the interlayer cation-N distance, see \textbf{Table~\ref{tab:interlayer}}. 

\begin{table}[ht!]
    \centering
    \begin{tabular}{c c c c c}
        Configuration & \multicolumn{4}{c}{Interlayer cation-N distance (\AA)}\\
        & cation 1 & cation 2 & cation 3 & cation 4 \\
        \hline
        InN & 2.19 & & &\\
        BN & 1.58 & & &\\
        2 B z-line & 1.53 & 1.54 & & \\
        3 B z-line & 2.17 & 2.17 & 2.17 & \\
        4 B z-line & 1.67 & 3.23 & 3.14 & 2.953 \\
        \hline
    \end{tabular}
    \caption{Interlayer cation-N bonds for BN, InN and different "z-line" configurations. Note that distance quoted is an In-N bond for InN and a B-N bond for all other systems. Cation numbers are such that the bottom of the "z-line" is cation 1.}
    \label{tab:interlayer}
\end{table}
In InN and BN this distance is the same throughout the crystal. In the 2 B case, the B-N distances are very close to those in pure wz BN, indicating thus a significant local deviation from the InN crystal structure. Furthermore, the resulting local strain effects are also expected to be accompanied by local polarization field effects which can give rise to carrier localization. This is confirmed by the charge densities shown in Figure~\ref{fig:2B-zline}, which indicate strong localization effects in the valence band (both in VBM but also the lower lying valence states, e.g. VBM-1); the charge densities of the conduction band states are affected to a much lesser extend related to the in general lower effective electron mass when compared to hole effective masses in III-N systems.~\cite{araujo13, de11}. For the 4 B atom case, Table~\ref{tab:interlayer} shows that the anion-cation interlayer spacing is also strongly impacted by the presence of B atoms, again indicating that the lattice structure is noticeably changed. For one B atom in the 4 B "z-line" configuration the interlayer B-N distance is much shorter compared to the equivalent layer spacing in an ideal InN lattice and only slightly longer when compared to BN. The remaining B-N distances in the lattice are not only longer than the interlayer spacing of pure wz BN but also exceed the equivalent In-N distances in pristine InN. This increased distance is in contrast to what we observed for the "line" configurations and suggests that the boron atoms behave differently when situated along the $c$-axis. Additionally, the increased distance may be large enough to prevent In-N bond formation in some areas of the crystal, which could have an effect on the quality of any experimentally grown (B,In)N. We note that to determine equilibrium positions and lattice parameters in our DFT cell, the pressure on the simulation cell is minimized. As such, internal degrees of freedom (e.g. bond lengths and angles) are optimized to achieve this, which may also mean that if layer distances are smaller in one region, layer spacing in other regions of the cell can be larger than e.g. in InN to compensate. Further, Figure~\ref{fig:BG-lat} (b) shows that for the 4 B "z-line" configuration the system with the smallest pressure on the cell leads to a supercell with a $c$ lattice constant that exhibits a larger deviation from a Vegard approximation when compared to other configurations. From the resulting conduction and valence band charge densities for the 4 B atom "z-line" case, Figure~\ref{fig:4B-zline}, we find carrier localization in the valence band (VBM and VBM-1) similar to the 2 B atom case in the "z-line" configuration; the conduction band states are only very weakly impacted by the presence of B atoms. Therefore one may expect that these carrier localization effects in the valence band lead to a shrinking band gap in such a (B,In)N alloy, and ultimately closing the small band gap that existed in pure InN. As for the "line" configurations, this is further supported by the defect bands shown in the densities of state, see Supporting Information. 

\textbf{\begin{figure}[ht!]
    \centering
    \includegraphics[width=\linewidth,keepaspectratio]{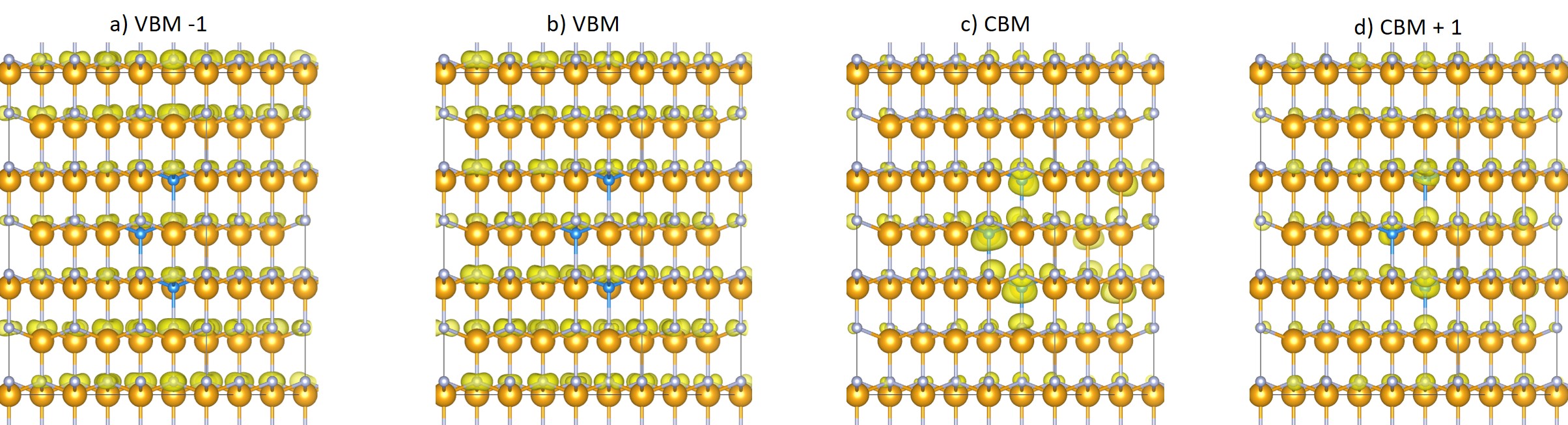}
    \caption{Band decomposed charge densities for a 3 B atoms in a vertical "z-line" in InN. See previous figure caption for information on color coding.}
    \label{fig:3B-zline}
\end{figure}}

As stated above, the 3 B atom "z-line" configuration, however, behaves differently to the equivalent 2 B and 4 B configurations. Firstly, the three interlayer B-N distances are very close to the distance expected in pristine InN, see Table~\ref{tab:interlayer}. This already suggests that the local environment, at least in terms of bond lengths and angles, is only weakly impacted by the presence of B atoms. Also, in contrast to the 2 B and 4 B atom "z-line" configuration, the valence band states VBM and VBM-1 give very little indications of carrier localization and can basically be described by delocalized states as shown in Figure~\ref{fig:3B-zline}. The conduction band states CBM and CBM+1 are also only slightly impacted by the presence of B atoms in the cell, see again Figure~\ref{fig:3B-zline}; the DOS shows no defect-related mid-gap states. All this behavior is consistent with the finding that 3 B atoms in a "z-line" configuration only slightly affect the InN band gap (i.e. increase the band gap by ca. 11\% or 0.07 eV as shown in Figure~\ref{fig:BG-lat} (a)).

Overall, these observations suggest columnar growth of B in III-N can have a significant impact on the electronic structure of B containing InN alloys. Future studies may focus on this question further as our current investigation indicates that if growth results in integer multiples of BN unit cells stacked along the wurtzite $c$-axis (2 B "z-line" = one unit cell, 4 B "z-line" = two unit cells) the band gap of (B,In)N alloys will be significantly reduced. This should be contrasted in the case of 3 B atom "z-line" configurations (half integer multiple of BN unit cells along the wurtzite $c$-axis), where the band gap is largely unaffected. It would be interesting to see if a 5 B atom "z-line" configuration again leads to a weakly impacted band gap or an even more dramatically increased band gap when compared to the InN band gap. However, such a study would require a larger supercell than the one used here and is beyond the scope of the present study.

\section{Conclusion}
We have presented first-principles electronic structure calculations for (B,In)N alloys with boron concentrations ranging from 2\% to 7\%. Special attention has been paid to the impact of the alloy microstructure, and in particular boron atom clustering, on the electronic structure of (B,In)N systems. While we find that the lattice constants of these alloys are well approximated by a Vegard interpolation, the electronic structure strongly varies with the local alloy microstructure. If boron atoms are dispersed throughout the (B,In)N alloy, thus not sharing nitrogen atoms, the band gap in the studied alloy content range is not significantly affected and remains close to that of pure InN. However, our calculations also reveal that in the case of boron atom clustering, when boron atoms share nitrogen atoms, the band gap may be significantly impacted and in most cases reduced. Here we find that local arrangement of boron atoms plays an important role (clustering within the $c$-plane or along the $c$-axis). In general our calculations suggest that boron atom clustering reduces the band gap, and in the case of (B,In)N may lead to a metallic state. We note that as for (B,Ga)N alloys, the large band gap reduction in (B,In)N is driven by carrier localization effects, which occur mainly in the valence band. 

In combination with our previous study on (B,Ga)N alloys, this suggests that if boron atom clustering can be avoided in (B,In,Ga)N alloys, one may control the lattice constant of the alloy by varying the boron content in the system, while the band gap is mainly determined by the In content. This finding is promising for achieving efficient light emission in the red spectral region based on (B,In,Ga)N alloys. 
However, given that (In,Ga)N alloys are already prone to carrier localization, these effects may be further enhanced when boron atoms cluster in (B,In,Ga)N alloys. Such clustering may in turn lead to a strong reduction in band gap for these alloys even with small fractions of boron. Further studies are required to understand the impact of alloy disorder and boron atom clustering on the electronic and optical properties of emerging, complex III-N alloys such as (B,In,Ga)N. An important aspect will also be to discover how the local alloy composition could be controlled during the growth process. Therefore, joint experimental and theoretical studies will be essential to gain insight into these questions.


\medskip
\section*{Data Availability} 
VASP inputs for the alloy configurations studied here and in our previous work~\cite{BGaN} are available on GitHub: \url{https://github.com/clnies/Research-Data/tree/main/BGaN}.

\medskip
\section*{Acknowledgements} 
This work was supported by Science Foundation Ireland, Grant Number 21/FFP-A/9014 and 12/RC/2276\_P2.

\clearpage

%
\bibliographystyle{MSP}
\bibliography{ref.bib}

\clearpage
\includepdf[pages={1-3}]{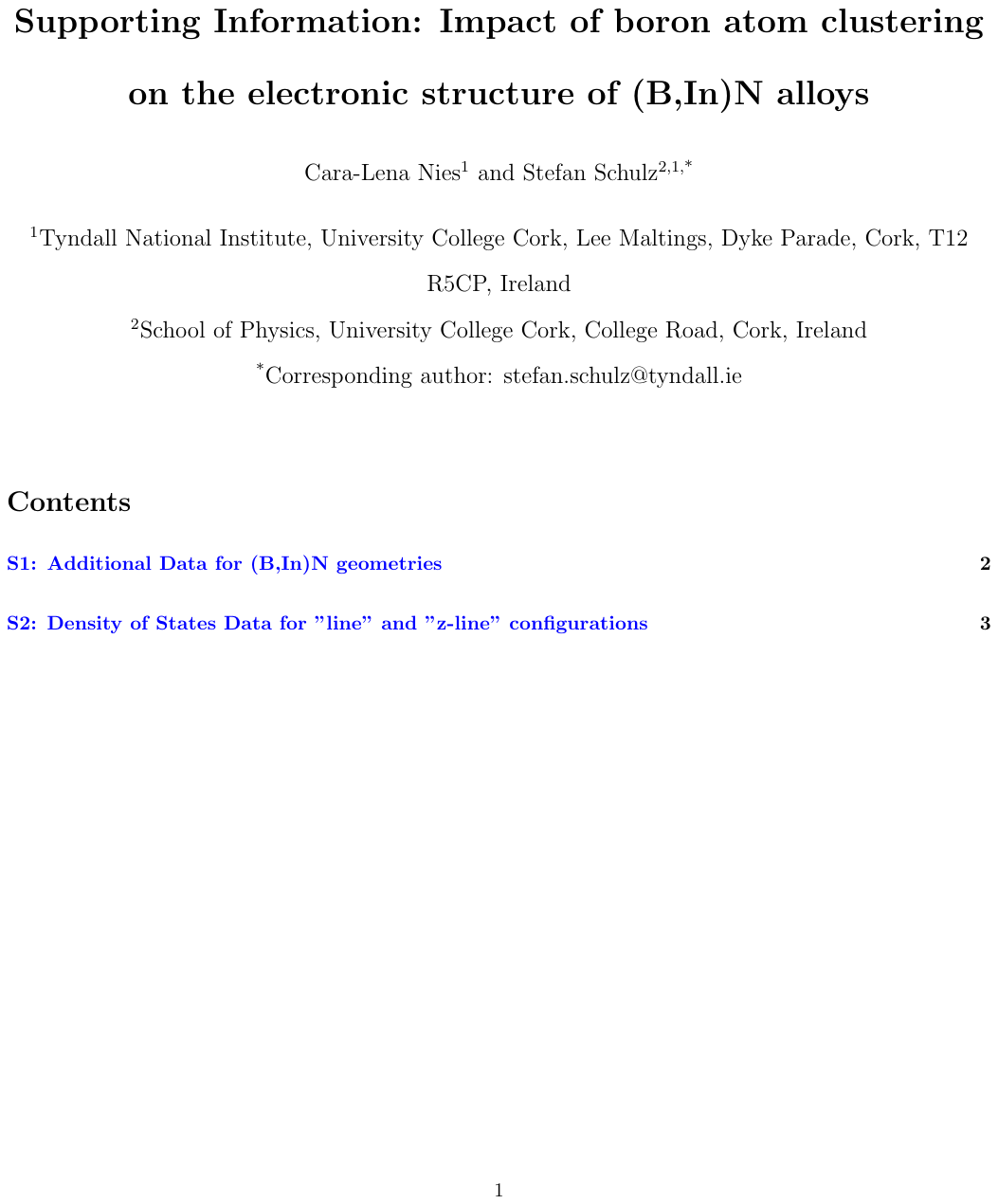}



\end{document}